\documentclass{article}
\usepackage{spconf,amsmath,amssymb,graphicx,booktabs,array,capt-of,etoolbox,xurl,placeins}
\usepackage[hidelinks]{hyperref}
\AtBeginEnvironment{thebibliography}{\setlength{\itemsep}{0pt}\setlength{\parsep}{0pt}\setlength{\parskip}{0pt}}
\hypersetup{pdftitle={NeuMark-Native: Robust Text-to-Speech-Native Watermarking through Full Utilization of Neural Audio Codec Latent Space},pdfauthor={Annan Wu, Wen-Chin Huang, Tomoki Toda}}
\newcommand{\wm}{\mathrm{wm}}
\newcommand{\Loss}{\mathcal{L}}
\newcommand{\xw}{x_{\wm}}

\newcommand{\cosim}{\operatorname{cos}}
\newcommand{\tablefont}{}
\makeatletter
\renewcommand\section{\@startsection{section}{1}{\z@}%
  {-3.5ex}{2.3ex}{\normalfont\Large\bfseries}}
\renewcommand\subsection{\@startsection{subsection}{2}{\z@}%
  {-3.0ex}{1.5ex}{\normalfont\large\bfseries}}
\makeatother
\title{NeuMark-Native: Robust Text-to-Speech-Native Watermarking through Full Utilization of Neural Audio Codec Latent Space}
\name{Annan Wu, Wen-Chin Huang, Tomoki Toda}
\address{Nagoya University, Japan}

\begin{document}
\ninept
\setlength{\parskip}{0pt}
\maketitle
\begin{abstract}
Speech watermarking offers proactive traceability for synthetic speech, yet most existing models operate only after text-to-speech (TTS) synthesis by adding a watermark perturbation to the generated waveform. This post-hoc design leaves watermarking as an external step that can be omitted or bypassed and restricts the watermark to a shallow waveform representation. We propose NeuMark-Native, a TTS-native watermarking framework for neural codec-based synthesis. It embeds payload information into every generated codec-latent layer before waveform decoding, improving watermark persistence under downstream digital signal processing (DSP) and neural codec resynthesis. NeuMark-Native keeps the pretrained TTS model and the neural codec frozen, while optimizing only the watermark modules on generated codec tokens. Experiments on two corpora under 11 DSP attacks and 9 neural-codec attacks demonstrate robust watermark detection while preserving naturalness, intelligibility, and speech quality close to synthetic speech.
\end{abstract}

\section{Introduction}
Recent neural codec-based text-to-speech models, including VALL-E~2~\cite{chen2024valle2}, MaskGCT~\cite{wang2024maskgct}, and CosyVoice~3~\cite{du2025cosyvoice3}, generate increasingly realistic synthetic speech, creating a growing need for reliable provenance. Speech watermarking provides proactive traceability by embedding recoverable information into synthetic speech. Research on speech watermarking has progressed from signal-processing-based methods using transform coding and auditory masking~\cite{ruiz2000speechwatermark} to neural methods that learn watermark embedding and recovery.

\begin{figure}[t]
\centering
\includegraphics[width=\columnwidth]{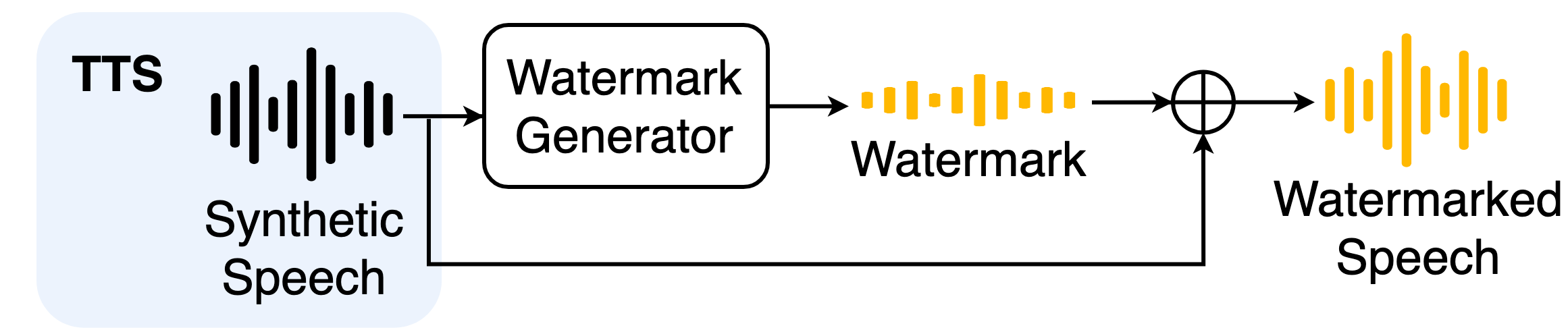}
\caption{Post-hoc watermarking applied after TTS synthesis.}
\label{fig:waveform-post}
\end{figure}

Existing neural speech watermarking methods commonly apply the watermark only after waveform generation~\cite{chen2023wavmark,sanroman2024audioseal,singh2024silentcipher,li2024ideaw,liu2025xattnmark}. Such post-hoc methods add a learned waveform perturbation to synthetic speech, as illustrated in Fig.~\ref{fig:waveform-post}. Although convenient and model agnostic, this interface confines the watermark to the final waveform. As a separate post-processing step, watermarking can be omitted or bypassed. Latent-Mark~\cite{chen2026latentmark} further shows that neural codec resynthesis can largely remove conventional waveform watermarks despite their robustness to ordinary digital signal processing (DSP) operations.

\begin{figure}[t]
\centering
\includegraphics[width=0.84\columnwidth]{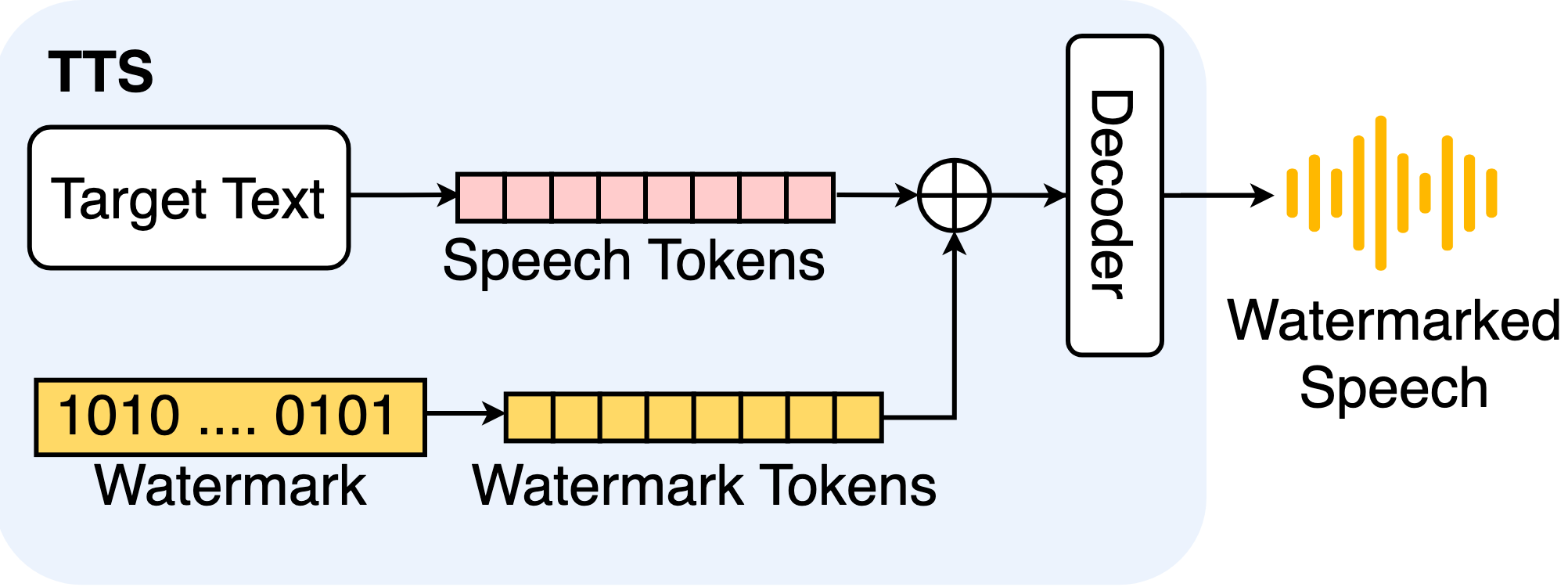}
\caption{TTS-native watermarking applied during TTS synthesis.}
\label{fig:tts-native-inference}
\end{figure}

Making watermarking native to TTS generation turns traceability into part of synthesis. TraceableSpeech~\cite{zhou2024traceablespeech} follows the TTS-native synthesis path illustrated in Fig.~\ref{fig:tts-native-inference}, embedding a watermark into one continuous codec representation and recovering it with a ResNet extractor. However, it first jointly trains the watermark model and speech codec, and then trains VALL-E~\cite{wang2023valle} on the resulting codec representations; any modification to the watermark model therefore requires VALL-E to be retrained from scratch. It also incurs a speech-quality cost and leaves robustness to neural codec resynthesis unaddressed.

\begin{figure}[t]
\centering
\includegraphics[width=0.98\columnwidth]{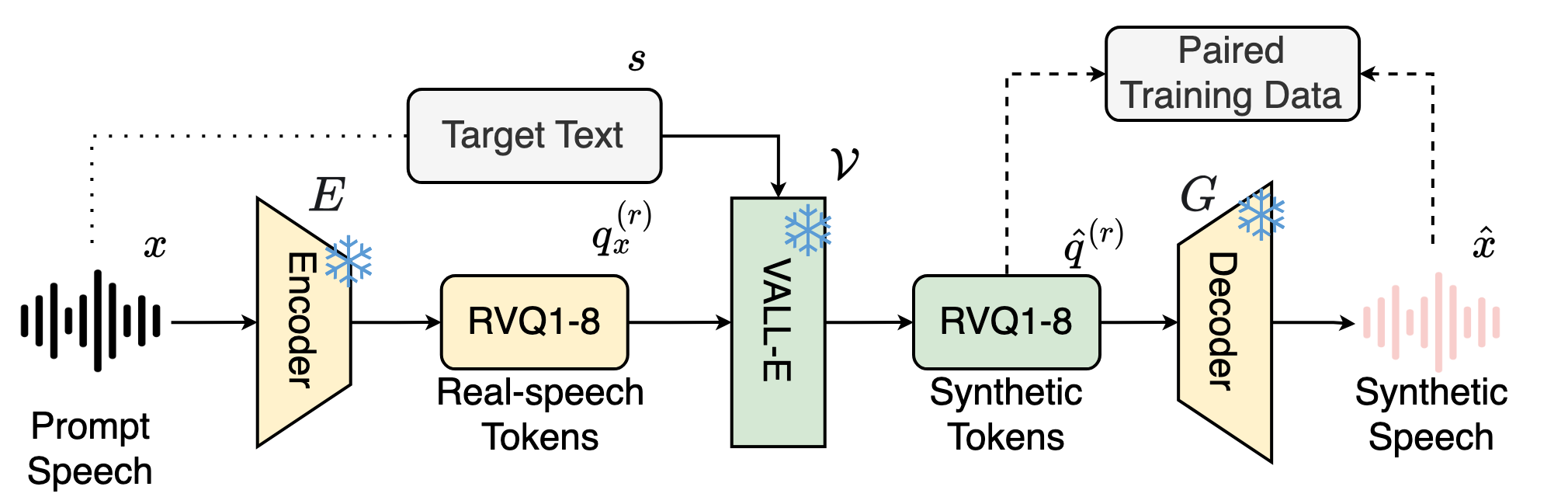}
\caption{Offline preparation of paired TTS-generated codec tokens and synthetic speech.}
\label{fig:prepare}
\end{figure}

\begin{figure*}[t]
\centering
\includegraphics[width=0.85\textwidth]{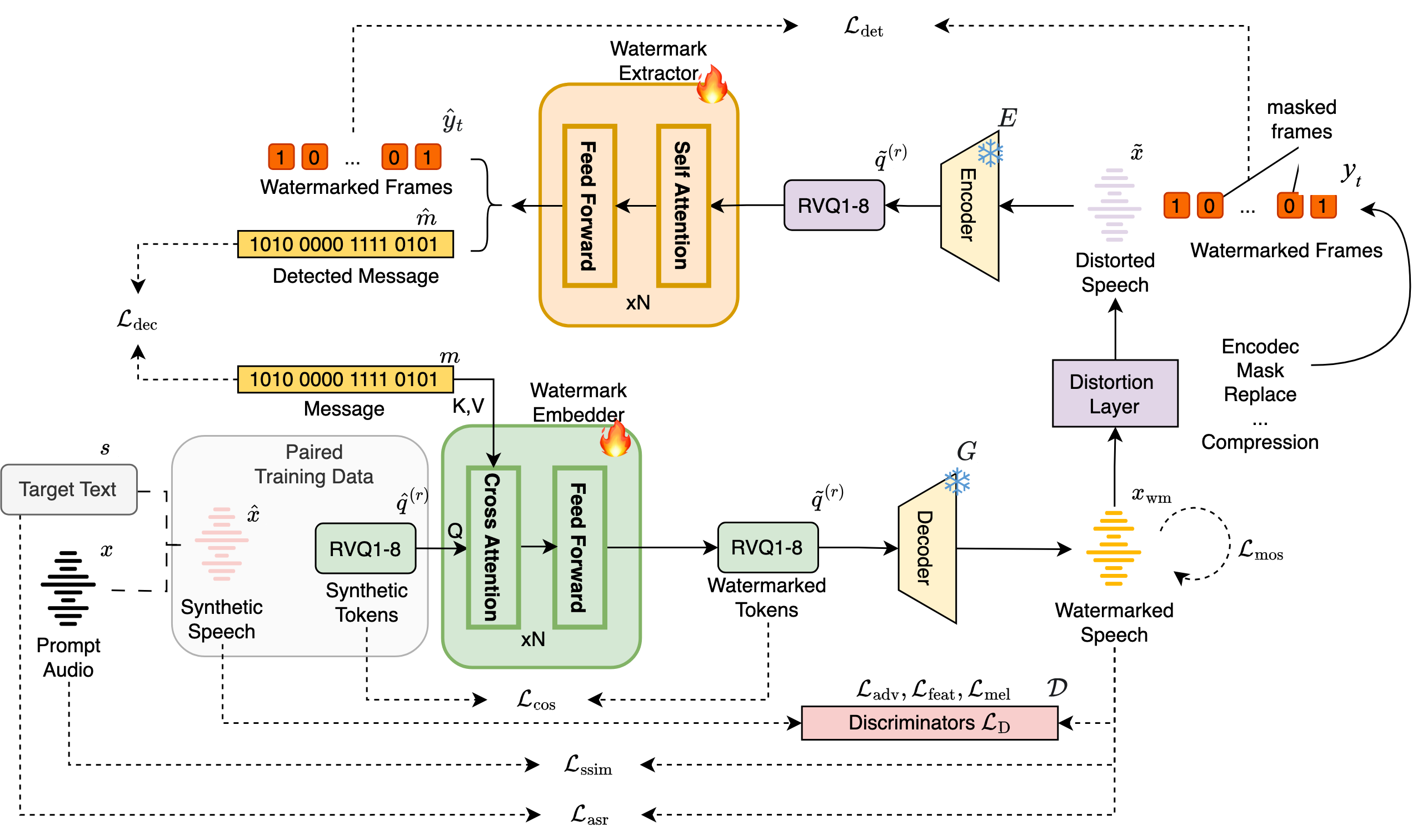}
\caption{NeuMark-Native training on the prepared TTS-generated codec tokens.}
\label{fig:pipeline}
\end{figure*}

To improve robustness to neural codec resynthesis without retraining the speech codec, we previously proposed NeuMark~\cite{wu2026neumark}, which keeps the speech codec frozen and optimizes only its Transformer watermark modules. Through cross-attention, these modules apply message-conditioned residual updates to every residual vector quantization (RVQ) layer, distributing watermark evidence across codec-aligned latent representations to keep the watermark detectable after resynthesis. However, because its carriers are encoded from real speech, watermarking remains separate from TTS generation.

In this paper, we propose NeuMark-Native, a TTS-native framework that extends NeuMark's per-layer codec-latent watermarking to TTS-generated tokens while retaining its robustness to neural codec resynthesis and frozen-backbone training strategy. A pretrained TTS model pre-synthesizes target codec tokens from each training transcript and a same-speaker prompt, exposing the watermark modules to TTS-specific characteristics while the TTS model and speech codec remain frozen. NeuMark-Native produces watermarked synthetic speech with quality competitive with waveform-domain methods such as AudioSeal~\cite{sanroman2024audioseal} and WavMark~\cite{chen2023wavmark}. Our contributions are:

\begin{list}{\textbullet}{%
\setlength{\leftmargin}{1.3em}%
\setlength{\labelwidth}{0.5em}%
\setlength{\labelsep}{0.5em}%
\setlength{\itemindent}{0pt}%
\setlength{\itemsep}{0pt}%
\setlength{\parsep}{0pt}%
\setlength{\topsep}{2pt}}
\item We propose NeuMark-Native, a TTS-native framework that adapts its watermark modules on pre-synthesized codec tokens carrying TTS-specific characteristics and distributions. It embeds payloads natively within codec-based synthesis using a loss design tailored to zero-shot voice cloning.
\item We evaluate NeuMark-Native on two corpora under 11 DSP and 9 neural-codec attacks. It substantially outperforms competing methods in codec-resynthesis robustness while achieving speech quality comparable to waveform-domain models across various evaluation metrics. NeuMark-Native source code is available at {\small\urlstyle{same}\url{https://github.com/goannan/NeuMark-Native}}.
\end{list}

\section{Proposed Method}
\subsection{Offline generated-token preparation}
VALL-E~\cite{wang2023valle} is a neural codec language model for zero-shot TTS. It takes target text and a speech prompt as input and predicts the codec tokens of target speech. The VALL-E model used here was pretrained to predict SpeechTokenizer~\cite{zhang2023speechtokenizer} tokens. As shown in Fig.~\ref{fig:prepare}, its input--output relation is
\begin{equation}
\begin{aligned}
\mathbf q_x &= E(x)=\{q_x^{(r)}\}_{r=1}^{R},\\
\hat{\mathbf q} &= \mathcal V(s,\mathbf q_x)=\{\hat q^{(r)}\}_{r=1}^{R},\\
\hat{x} &= G(\hat{\mathbf q}).
\end{aligned}
\label{eq:valle-pipeline}
\end{equation}
Here $s$ is the target text; $x$ is a same-speaker prompt-speech waveform; $E$ and $G$ are the frozen SpeechTokenizer encoder and decoder; $\mathbf q_x$ is the set of prompt tokens; $\mathcal V$ is the frozen VALL-E model; $\hat{\mathbf q}$ is the set of predicted target tokens; $q_x^{(r)}$ and $\hat q^{(r)}$ denote the prompt and target streams at RVQ layer $r$, respectively; $R$ is the number of RVQ layers; and $\hat{x}$ is the resulting synthetic speech.

For offline adaptation, $s$ is the transcript of a training utterance and $x$ is a prompt from the same speaker. As shown in Fig.~\ref{fig:prepare}, each pre-synthesis run stores the predicted tokens $\hat{\mathbf q}$ together with their decoded synthetic speech $\hat{x}$ as a paired training sample. The TTS-generated tokens, rather than tokens encoded from the target recording, serve as the watermark carrier, while $\hat{x}$ provides the sample-aligned reference. This exposes the trainable modules to the token characteristics encountered during synthesis.

\subsection{TTS-native training objective}
Figure~\ref{fig:pipeline} shows how the cached TTS-generated tokens $\hat{\mathbf q}$ become the actual watermark carrier. For each random 16-bit message $m$, the carrier supplies the cross-attention queries $Q$, while $m$ supplies keys and values $K,V$; $N$ denotes the number of stacked Transformer blocks. The embedder modifies every RVQ stream and produces watermarked tokens $\widetilde{\mathbf q}$, which the frozen decoder $G$ converts into watermarked synthetic speech $\xw$. In parallel, decoding the unchanged $\hat{\mathbf q}$ produces the paired reference $\hat{x}$. The distortion layer transforms $\xw$ into $\widetilde{x}$ and provides frame labels $y_t$; the frozen encoder $E$ maps $\widetilde{x}$ to distorted tokens $\widetilde q^{(r)}$; and the extractor outputs predictions $\hat y_t$ and the recovered message $\hat m$. Flame and snowflake symbols denote trainable and frozen modules. Thus, only the watermark modules learn from the difference between the paired branches.

NeuMark-Native retains six NeuMark objectives~\cite{wu2026neumark}. The decoding, detection, latent-cosine, adversarial, feature-matching, and mel losses are denoted by $\Loss_{\mathrm{dec}}$, $\Loss_{\mathrm{det}}$, $\Loss_{\mathrm{cos}}$, $\Loss_{\mathrm{adv}}$, $\Loss_{\mathrm{feat}}$, and $\Loss_{\mathrm{mel}}$, respectively. The paired reference branch provides a sample-aligned anchor that prevents $\xw$ from drifting in content or acoustics. We further introduce three speech-aware objectives.

The naturalness loss is
\begin{equation}
\Loss_{\mathrm{mos}}=-U(\xw),
\label{eq:mos}
\end{equation}
where $U$ is the frozen UTMOS~\cite{saeki2022utmos} predictor, which maps watermarked synthetic speech $\xw$ to a predicted mean-opinion score. Minimizing $\Loss_{\mathrm{mos}}$ therefore encourages a higher naturalness score.

The speaker-similarity (SSIM) loss is
\begin{equation}
\Loss_{\mathrm{ssim}}=1-\cosim(V(\xw),V(x)),
\label{eq:ssim}
\end{equation}
where $V$ is an ECAPA-TDNN speaker-verification model using frozen WavLM-Large~\cite{chen2022wavlm} features and its official fine-tuned checkpoint,\footnote{\url{https://github.com/microsoft/UniSpeech/tree/main/downstreams/speaker_verification}} and $\cosim$ denotes cosine similarity. Minimizing $\Loss_{\mathrm{ssim}}$ preserves the prompt speaker's identity in $\xw$.

The ASR loss is
\begin{equation}
\Loss_{\mathrm{asr}}=\operatorname{CTC}\!\left(\log\operatorname{softmax}(B(\xw)),\tau(s)\right),
\label{eq:asr}
\end{equation}
where $B$ is the frozen wav2vec~2.0 Base (960 h) model~\cite{baevski2020wav2vec2}\footnote{\url{https://docs.pytorch.org/audio/stable/pipelines}} and maps $\xw$ to framewise character logits. The tokenizer $\tau$ converts target text $s$ into a character sequence. Connectionist temporal classification (CTC)~\cite{graves2006ctc} sums the probabilities of all valid frame-level paths corresponding to $\tau(s)$ and returns their negative log-likelihood. Minimizing $\Loss_{\mathrm{asr}}$ encourages $\xw$ to preserve the target linguistic content without frame-aligned labels.

Our joint adaptation objective is
\begin{equation}
\begin{aligned}
\Loss={}&5\Loss_{\mathrm{dec}}+\Loss_{\mathrm{det}}+2\Loss_{\mathrm{cos}}
+\Loss_{\mathrm{adv}}+2\Loss_{\mathrm{feat}}\\
&+0.25\Loss_{\mathrm{mel}}+0.5\Loss_{\mathrm{mos}}
+\Loss_{\mathrm{ssim}}+0.5\Loss_{\mathrm{asr}}.
\end{aligned}
\label{eq:total}
\end{equation}
For adversarial training, $\mathcal D$ contains multi-period, multi-scale, and multi-scale STFT discriminators (MPD, MSD, and MS-STFTD). Under the least-squares objective, discriminator $D\in\mathcal D$ with $K_D$ outputs is updated by
\begin{equation}
\Loss_D=\frac{1}{|\mathcal D|}\sum_{D\in\mathcal D}\frac{1}{K_D}\sum_{k=1}^{K_D}
\left[\mathbb E(D_k(\hat{x})-1)^2+\mathbb E D_k(\xw)^2\right].
\label{eq:disc}
\end{equation}
The watermark modules are optimized by Eq.~\eqref{eq:total} with the discriminators frozen, whereas $\mathcal D$ is updated separately by Eq.~\eqref{eq:disc} using a detached $\xw$. All TTS, codec, and speech-assessment networks remain frozen.

We evaluate two NeuMark-Native variants. \emph{NeuMark-Native (direct)} applies the NeuMark watermark modules pretrained using original codec tokens extracted from natural speech directly to TTS-generated tokens. \emph{NeuMark-Native (adapted)} initializes from the same checkpoint and fine-tunes the modules on the paired training data in Fig.~\ref{fig:prepare} using Eq.~\eqref{eq:total}. Both variants keep the TTS model and speech codec frozen and use the same watermark-embedding procedure during synthesis.

\section{Experimental Evaluation}
\subsection{Setup}
NeuMark-Native used the three LibriTTS~\cite{zen2019libritts} training subsets: clean-100, clean-360, and other-500. Together they contained approximately 555~hours before duration filtering. Filtering to 0.5--10.5~sec yielded approximately 157k cuts, which were preprocessed as in Sec.~2.1. Adaptation ran for ten epochs on one NVIDIA H100 GPU and took approximately 69~hours. Each example received a random 16-bit message and one augmentation sampled from identity or 11 DSP attacks following AudioCraft:\footnote{\url{https://github.com/facebookresearch/audiocraft}} Gaussian noise, pink noise, low-pass filtering, high-pass filtering, band-pass filtering, echo, smoothing, resampling, gain increase, gain decrease, and speed change. The training pool also contained EnCodec~\cite{defossez2022encodec} resynthesis at 3, 6, and 12~kbps to adapt the model to neural-codec attacks across different bandwidths, together with codec resynthesis combined with frame masking or replacement; negatives alternated between recordings and synthetic speech.

Evaluation was conducted as zero-shot text-to-speech (TTS) voice cloning on both corpora. It used 2,078 duration-filtered utterances from LibriTTS test-clean as the in-corpus set and 1,088 prompt--text pairs from the English subset of Seed-TTS-Eval~\cite{anastassiou2024seedtts} (SeedTTS-en) to assess cross-corpus generalization to unseen texts and speakers. Each test item first followed Fig.~\ref{fig:prepare}: from test text $s$ and a same-speaker prompt $x$, VALL-E predicted target speech tokens $\hat{\mathbf q}$, which the frozen decoder converted into synthetic speech $\hat{x}$.

NeuMark-Native and TraceableSpeech performed watermark embedding within TTS generation before waveform decoding, thereby producing watermarked synthetic speech directly. By contrast, WavMark and AudioSeal operated after TTS synthesis by adding watermark perturbations to the synthetic speech $\hat{x}$ produced in Fig.~\ref{fig:prepare}. The resulting watermarked synthetic speech from every method was evaluated under the same DSP and neural-codec conditions.

The evaluation included one identity condition and the same 11 DSP attacks. The 9 codec attacks used EnCodec at 3/6/12~kbps, DAC~\cite{kumar2023dac} at 6/8/24~kbps, and SNAC~\cite{siuzdak2024snac} at 0.98/1.9/2.6~kbps. DAC and SNAC were unseen during training. Group means formed the 12/9-weighted overall score.

\begin{table*}[!t]
\centering
\begin{minipage}{0.96\textwidth}
\centering
\captionof{table}{Robustness under DSP and codec conditions. AUC: detection ROC-AUC; TPR$_{.1}$: TPR at FPR $\leq0.1\%$; Bit: payload accuracy. DSP, Codec, and Overall report means over their respective conditions.}
\label{tab:robustness}
\tablefont\setlength{\tabcolsep}{6pt}
\begin{tabular*}{\linewidth}{@{\extracolsep{\fill}}llrrrrrr@{}}
\toprule
& & \multicolumn{3}{c}{LibriTTS test-clean} & \multicolumn{3}{c}{SeedTTS-en} \\
\cmidrule(lr){3-5}\cmidrule(l){6-8}
Method & Conditions & AUC$\uparrow$ & TPR$_{.1}\uparrow$ & Bit$\uparrow$ & AUC$\uparrow$ & TPR$_{.1}\uparrow$ & Bit$\uparrow$ \\
\midrule
WavMark~\cite{chen2023wavmark} & DSP & 0.993 & 0.931 & 0.954 & 0.989 & \textbf{0.925} & 0.948 \\
 & Codec & 0.570 & 0.095 & 0.542 & 0.572 & 0.092 & 0.543 \\
 & Overall & 0.812 & 0.572 & 0.778 & 0.810 & 0.568 & 0.775 \\
\addlinespace[2pt]
AudioSeal~\cite{sanroman2024audioseal} & DSP & 0.970 & 0.920 & 0.946 & 0.971 & 0.920 & 0.944 \\
 & Codec & 0.813 & 0.445 & 0.638 & 0.806 & 0.422 & 0.645 \\
 & Overall & 0.903 & 0.717 & 0.814 & 0.900 & 0.707 & 0.816 \\
\addlinespace[2pt]
TraceableSpeech~\cite{zhou2024traceablespeech} & DSP & 0.987 & 0.921 & 0.958 & 0.984 & 0.921 & 0.955 \\
 & Codec & 0.660 & 0.088 & 0.605 & 0.697 & 0.090 & 0.610 \\
 & Overall & 0.847 & 0.564 & 0.807 & 0.861 & 0.565 & 0.807 \\
\addlinespace[2pt]
NeuMark-Native (direct) & DSP & 0.999 & 0.948 & 0.974 & \textbf{0.998} & 0.885 & \textbf{0.973} \\
 & Codec & 0.942 & 0.521 & \textbf{0.861} & 0.940 & 0.481 & \textbf{0.885} \\
 & Overall & 0.975 & 0.765 & \textbf{0.926} & 0.973 & 0.712 & \textbf{0.936} \\
\addlinespace[2pt]
NeuMark-Native (adapted) & DSP & \textbf{1.000} & \textbf{0.961} & \textbf{0.982} & \textbf{0.998} & 0.914 & 0.964 \\
 & Codec & \textbf{0.955} & \textbf{0.605} & 0.842 & \textbf{0.949} & \textbf{0.490} & 0.830 \\
 & Overall & \textbf{0.981} & \textbf{0.808} & 0.922 & \textbf{0.977} & \textbf{0.733} & 0.907 \\
\bottomrule
\end{tabular*}
\end{minipage}

\end{table*}
\begin{table*}[!t]
\centering
\begin{minipage}{0.96\textwidth}
\centering
\captionof{table}{Speech quality. Synthetic speech is the common unwatermarked carrier; PESQ~\cite{pesq} and STOI~\cite{taal2011stoi} use it as reference.}
\label{tab:quality}
\tablefont\setlength{\tabcolsep}{2pt}
\begin{tabular*}{\linewidth}{@{\extracolsep{\fill}}l*{10}{c}@{}}
\toprule
& \multicolumn{5}{c}{LibriTTS test-clean} & \multicolumn{5}{c}{SeedTTS-en} \\
\cmidrule(lr){2-6}\cmidrule(l){7-11}
Method & UTMOS$\uparrow$ & SSIM$\uparrow$ & WER(\%)$\downarrow$ & PESQ$\uparrow$ & STOI$\uparrow$ & UTMOS$\uparrow$ & SSIM$\uparrow$ & WER(\%)$\downarrow$ & PESQ$\uparrow$ & STOI$\uparrow$ \\
\midrule
Synthetic speech & 3.664 & 0.425 & 35.8 & --- & --- & 3.276 & 0.420 & 36.7 & --- & --- \\
\midrule
WavMark & 3.632 & 0.421 & \textbf{35.7} & 4.159 & 0.996 & 3.224 & 0.414 & \textbf{36.8} & 4.081 & 0.988 \\
AudioSeal & \textbf{3.651} & \textbf{0.425} & 35.8 & 4.416 & \textbf{0.997} & \textbf{3.259} & \textbf{0.417} & \textbf{36.8} & \textbf{4.427} & 0.990 \\
TraceableSpeech & 2.198 & 0.174 & 41.3 & 3.048 & 0.925 & 2.109 & 0.182 & 47.1 & 2.463 & 0.895 \\
NeuMark-Native (direct) & 3.638 & 0.410 & \textbf{35.7} & 4.360 & 0.990 & 3.244 & 0.403 & 36.9 & 4.161 & 0.979 \\
NeuMark-Native (adapted) & 3.630 & 0.413 & \textbf{35.7} & 4.318 & 0.991 & 3.238 & 0.413 & 37.0 & 4.212 & 0.985 \\
\midrule
\emph{w/o} native losses & 3.648 & 0.420 & 35.8 & \textbf{4.443} & 0.995 & 3.254 & 0.414 & 37.0 & 4.382 & \textbf{0.992} \\
\emph{w/o} generated tokens & 3.613 & 0.413 & \textbf{35.7} & 4.278 & 0.989 & 3.228 & 0.407 & 36.9 & 4.058 & 0.978 \\
\bottomrule
\end{tabular*}
\end{minipage}

\end{table*}
\begin{table*}[!t]
\centering
\begin{minipage}{0.96\textwidth}
\centering
\captionof{table}{NeuMark-Native ablations. TTS-native losses are UTMOS, WavLM, and ASR objectives.}
\label{tab:ablation}
\tablefont\setlength{\tabcolsep}{3.5pt}
\begin{tabular*}{\linewidth}{@{\extracolsep{\fill}}lccrrrrrr@{}}
\toprule
& TTS-gen. & TTS-native & \multicolumn{3}{c}{LibriTTS test-clean} & \multicolumn{3}{c}{SeedTTS-en} \\
\cmidrule(lr){4-6}\cmidrule(l){7-9}
Configuration & tokens & losses & AUC$\uparrow$ & TPR$_{.1}\uparrow$ & Bit$\uparrow$ & AUC$\uparrow$ & TPR$_{.1}\uparrow$ & Bit$\uparrow$ \\
\midrule
NeuMark-Native (adapted) & Yes & Yes & \textbf{0.981} & 0.808 & 0.922 & \textbf{0.977} & 0.733 & 0.907 \\
\emph{w/o} native losses & Yes & No & 0.974 & 0.772 & 0.899 & 0.968 & 0.671 & 0.874 \\
\emph{w/o} generated tokens & No & Yes & 0.979 & \textbf{0.822} & 0.924 & 0.974 & \textbf{0.789} & 0.924 \\
NeuMark-Native (direct) & No & No & 0.975 & 0.765 & \textbf{0.926} & 0.973 & 0.712 & \textbf{0.936} \\
\bottomrule
\end{tabular*}
\end{minipage}

\end{table*}

\nopagebreak[4]
\subsection{Performance evaluation}
Table~\ref{tab:robustness} compares watermark robustness across the identity, DSP, and neural-codec conditions. NeuMark-Native (adapted) achieves the best overall detection performance on both LibriTTS test-clean and SeedTTS-en, while NeuMark-Native (direct) provides the highest overall payload-bit accuracy. Consistent results on both corpora demonstrate generalization to unseen texts and speakers.

Table~\ref{tab:quality} compares speech quality against the same synthetic speech. NeuMark-Native preserves naturalness, speaker similarity, intelligibility, and perceptual fidelity close to the unwatermarked reference, with quality competitive with the waveform-domain methods WavMark and AudioSeal. These results demonstrate that TTS-native watermarking can improve overall robustness without substantially degrading the generated speech.

\subsection{Ablation study}
The \emph{w/o native losses} variant removes the UTMOS, WavLM-similarity, and ASR objectives. The \emph{w/o generated tokens} variant replaces VALL-E-predicted tokens with SpeechTokenizer tokens encoded from real speech, as in NeuMark. Both retain the same initialization and NeuMark objectives.

Tables~\ref{tab:quality} and~\ref{tab:ablation} show complementary ablation strengths. Without TTS-native losses, the differences from AudioSeal are only 0.003/0.005 in UTMOS and 0.005/0.003 in SSIM on LibriTTS test-clean/SeedTTS-en, with the best LibriTTS PESQ and SeedTTS-en STOI. This confirms that NeuMark's reconstruction objectives provide a strong transparency anchor. Without generated tokens, low-FPR detection and payload recovery exceed the full adapted model despite modestly lower quality. Generated-token training therefore favors the overall robustness--quality balance, while real-speech tokens favor strict low-FPR detection and bit recovery; direct integration retains the highest codec-average payload accuracy.

\section{Conclusion}
NeuMark-Native integrates watermarking into neural codec TTS by adapting NeuMark on pre-synthesized TTS tokens and embedding the payload into every generated RVQ stream before decoding. Across two corpora, it provides the strongest overall detection and low-FPR performance, with particularly large gains under neural-codec resynthesis, while retaining naturalness, intelligibility, speaker similarity, and perceptual quality close to synthetic speech and competitive waveform-domain methods. The frozen-backbone design also permits transfer to compatible codec-based TTS models through watermark-only adaptation.

\section{Acknowledgment}
This work was supported in part by JSPS KAKENHI Grant Number 26H02530, and in part by the BRIDGE Program (R7-H05), implemented by the Cabinet Office, Government of Japan.

\bibliographystyle{IEEEbib}
\bibliography{references}
\end{document}